\documentclass[a4paper,10pt,twoside]{cpc-hepnp}

\usepackage{multicol}
\usepackage{graphicx}
\usepackage{booktabs}
\usepackage{amssymb,bm,mathrsfs,bbm,amscd}
\usepackage[tbtags]{amsmath}
\usepackage{lastpage}
\usepackage{CJK}

\begin{document}

\fancyhead[co]{\footnotesize LU Xin-Yu~ et al: Study of measuring methods on spatial resolution of a GEM imaging detector}

\footnotetext[0]{Received 14 March 2009}

\title{Study of measuring methods on spatial resolution of a GEM imaging detector\thanks{Supported by the Knowledge Innovation Program of the Chinese Academy of Sciences}}

\author{%
LU Xin-Yu$^{1,2;1)}$\email{lvxy@ihep.ac.cn}%
\quad FAN Rui-Rui$^{1}$
\quad CHEN Yuan-Bo$^{1,3}$\\
\quad OUYANG Qun$^{1,3}$
\quad LIU Rong-Guang$^{1,3}$
\quad LIU Peng$^{1}$
\quad QI Hui-Rong$^{1,3}$\\
\quad ZHANG Jian$^{1,3}$
\quad ZHAO Ping-Ping$^{1,3}$
\quad ZHAO Dong-Xu$^{1}$
\quad ZHAO Yu-Bin$^{1,3}$\\
\quad ZHANG Hong-Yu$^{1,3}$
\quad SHENG Hua-Yi$^{1,3}$
\quad DONG Li-Yuan$^{4}$
}
\maketitle

\address{%
$^1$ Institute of High Energy Physics, Chinese Academy of Sciences, Beijing 100049, China\\
$^2$ Graduate University of Chinese Academy of Sciences, Beijing 100049, China\\
$^3$ State Key Laboratory of Particle Detection and Electronics, Beijing 100049, China\\
$^4$ School of Nuclear Science and Technology, Lanzhou University, Lanzhou 730000, China
}

\begin{abstract}
In this paper, limitations of the common method measuring intrinsic spatial resolution of the GEM imaging detector are presented. Through theoretical analysis and experimental verification, we have improved the common method to avoid these limitations. Using these improved methods, more precise measurement of intrinsic spatial resolution are obtained.
\end{abstract}



\begin{multicols}{2}

\section{Introduction}

The GEM(Gas Electron Multiplier) is a typical Micro-Pattern gaseous detector, first invented in high energy physics, then applied in many other fields. We have constructed a 2-D imaging detector using triple-GEM for BRSF(Beijing Synchrotron Radiation Facility) with active area of $200mm\times200mm$, which is able to detect X-ray with high spatial resolution(Fig.~\ref{pic:gem}). The GEM consists of a thin, metal-coated polymer foil, etched by high density of holes. When potential is applied between electrodes up and down, radiation electrons drift into the holes, multiply and transfer to the other side. Each hole acts as an individual proportional amplifier. In our detector, three GEM foils are used to get much higher gain.

One important specification of an imaging detector is the intrinsic spatial resolution(resolution will be used for short later in this paper). There are several definitions of resolution according to different criteria, however, all of them are equivalent in fact. Take FWHM and the standard deviation $\sigma_0$ for instance: when the consideration function is the normal distribution, the relationship between standard deviation $\sigma_0$ and FWHM is $FWHM=2\sqrt{2ln2}\sigma_0$. In this paper, the standard deviation is chosen to be used as the resolution\cite{lab1, lab2}. Overall, resolution is to describe the resolving power of imaging detectors, so how to measure the resolution accurately is important.

\begin{center}
\includegraphics[width=80mm]{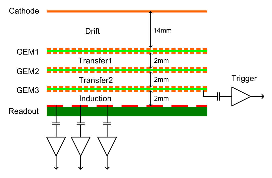}
\figcaption{\label{pic:gem}The schematic view of our GEM detector structure.}
\end{center}

\section{Analysis of the measurement of spatial resolution of imaging detectors}

\subsection{The common method to measure spatial resolution}

The common method to measure the resolution is usually to let beams collimated by a slit(sometimes by a hole or a blade)(Fig.~\ref{pic:slit}). The measurement result is always a distribution, of which standard deviation is $\sigma$. Then the resolution of the detector, of which standard deviation is $\sigma_0$, is obtained by Eq.~(\ref{eq:common}).

\begin{equation}\label{eq:common}
\sigma_0=\sqrt{{\sigma}^2-h^2}
\end{equation}
where $h$ is the width of the slit.

\begin{center}
\includegraphics[width=80mm]{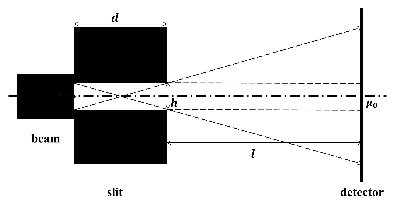}
\figcaption{\label{pic:slit}The schematic view of beams collimated by a slit. The X-ray beams go through the collimator, which is horizontal. The detector surface is vertical.}
\end{center}

We have done the measurement using this method in the condition: $d=20mm$, $h=0.2mm$, and $l=40mm$. The $\sigma$ of experimental data distribution is $1.33mm$. The resolution is calculated by Eq.~(\ref{eq:common}).

\begin{equation}\label{eq:commoncalculate}
\sigma{}_0=\sqrt{{1.33}^2-0.2^2}=1.3149mm
\end{equation}

Through a careful study, we have found limitations of this method and the error introduced in this method({Sec.~\ref{sec:limitation}}).

\subsection{Resolution of the detector, beam intensity distribution and experimental data distribution}\label{sec:relationship}

When we use the GEM detector for X-ray imaging of an object, the beam(after going through the collimator) intensity on surface of the detector is always a random variable, of which distribution is called {\bf beam intensity distribution}. The distribution of data measured by the detector is called {\bf experimental data distribution}. Due to the electrons diffusion in the detector and error existing in the measurement, the experimental data distribution can not reflect the resolution function accurately. Experimental data distributions are often the result of the resolution function that is modified by the beam intensity distribution. In fact, the experimental data random variable $Z$ is a superposition of the beam intensity random variable $X$ and detector resolution random variable $Y$ as shown in Eq.~(\ref{eq:experiment}).

\begin{equation}\label{eq:experiment}
Z=X+Y
\end{equation}

According to the central limit theorem, the resolution function usually follows the normal distribution. Hence, in the most general case, experimental data distributions are described by a convolution of a beam intensity distribution and the detector resolution function, where the p.d.f.(probability density function) of experimental data distributions is $g(x')$, that of  beam intensity distribution $f(x)$ and that of resolution function $r(x'-x)$ (Eq.~(\ref{eq:convolution})).

\begin{equation}\label{eq:convolution}
g(x')=\int_{\Omega{}_{x}}r(x'-x)f(x){\rm d} x
\end{equation}
$\Omega{}_{x}$ is the domain of the beam intensity random variable $X$. Its specific form is shown in Eq.~(\ref{eq:surface}) and Eq.~(\ref{eq:normalized}).

Hence, the resolution can be calculated by fitting the experimental data distribution with $g(x')$ ({Sec.~\ref{sec:method1}}) or calculating $r(x'-x)$ by deconvolution({Sec.~\ref{sec:method2}}).

\subsection{Limitations of the common method}\label{sec:limitation}

With different beam intensity distributions, different experimental data distributions are obtained accordingly, as shown in Table~\ref{tab:distribution}. From the table, only when the beam intensity distribution and the detector resolution function both follow the normal distribution, can the resolution be calculated by Eq.~(\ref{eq:common}). In many situations the beam intensity distribution does not follow normal distribution. In these conditions, if the width of the slit is smaller than tenth of the resolution(FWHM), the common method is still applicable according to GB/T 18989-200\footnote{Radionuclide imaging device performance and test rules for gamma camera}. This is the scope of application of the common method. To avoid this limitation, methods are improved to get more precise resolution. Experiments with GEM detectors\cite{lab3, lab4} have been done to confirm these improved methods({Sec.~\ref{sec:improved}}).

\end{multicols}
\begin{center}
\tabcaption{\label{tab:distribution} Experimental data distributions as convolution of resolution function and different beam intensity distributions.}
\footnotesize
\begin{tabular*}{\textwidth}{c@{\extracolsep{\fill}}cc}
\toprule
p.d.f. of resolution function&p.d.f. of beam intensity distribution&p.d.f. of experimental data distribution\\
\hline
$\frac{1}{\sqrt{2\pi}{\sigma_0}}e^{-\frac{(x'-x)^2}{2{\sigma_0}^2}}$&$f(x)=\delta(x-x_0)$&$g(x')=\frac{1}{\sqrt{2\pi}{\sigma_0}}e^{-\frac{(x'-x_0)^2}{2{\sigma_0}^2}}$\\
&(extremely narrow slit)&\vspace{3mm}\\
$\frac{1}{\sqrt{2\pi}{\sigma_0}}e^{-\frac{(x'-x)^2}{2{\sigma_0}^2}}$&$f(x)=\frac{1}{b-a}$&$g(x')=\frac{1}{b-a}(F(\frac{b-x'}{{\sigma_0}})-F(\frac{a-x'}{{\sigma_0}}))$*\\
&(uniform distribution: $a, b$ is the lower and upper limit)&\vspace{3mm}\\
$\frac{1}{\sqrt{2\pi}{\sigma_0}}e^{-\frac{(x'-x)^2}{2{\sigma_0}^2}}$&$f(x)=\frac{1}{\sqrt{2\pi}h}e^{-\frac{(x-x_0)^2}{2h^2}}$&$g(x')=\frac{1}{\sqrt{2\pi(h^2+{\sigma_0}^2)}}e^{-\frac{(x'-x_0)^2}{2(h^2+{\sigma_0}^2)}}$\\
&(normal distrubution)&\\
\bottomrule
\end{tabular*}
\end{center}
* $F(x)$ is cumulative distribution function of $f(x)$.
\vspace{1mm}
\begin{multicols}{2}

\section{Improved methods to measure the spatial resolution of imaging detectors}{\label{sec:improved}}

\subsection{Improved method~1: using a slit as the collimator with convolution fit}
{\label{sec:method1}}

As shown in Fig.~\ref{pic:slit}, the surface of detector, which is $d+l$ away from the beam source, is perpendicular to the beam line. So Fig.~\ref{pic:slit} actually shows the one dimension projection of the measurement frame. The beam source is a narrow strip source limited by the slit with a definite emission angle associated with the total domain $\Omega{}_{x}$ in Eq.~(\ref{eq:convolution}). The total effect is that the intensity along the X-axis on the surface of the detector is proportional to the range of the source from where the X-ray beam can reach the surface. Due to the effect of the collimator, the beam intensity distribution is divided into three parts. P.d.f. of the beam intensity distribution is as shown in Eq.~(\ref{eq:surface}).

\begin{equation}\label{eq:surface}
f(x)=\alpha \cdot \left\{ \begin{array}{ll}
h-\frac{d}{l}(x-\mu_0-\frac{h}{2})&x\in\big[\frac{h}{2}+\mu_0, (\frac{h}{2}+h\frac{l}{d})\\&+\mu_0\big]
\vspace{2mm}\\
h&x\in\big(-\frac{h}{2}+\mu_0, \frac{h}{2}+\mu_0\big)
\vspace{2mm}\\
h+\frac{d}{l}(x-\mu_0+\frac{h}{2})&x\in\big[-(\frac{h}{2}+h\frac{l}{d})+\mu_0, \\&-\frac{h}{2}+\mu_0\big]
\vspace{2mm}\\
\end{array}\right.
\end{equation}
where $\mu_0$ is the coordinate of the middle of the slit. $f(x)$ should be normalized.
\end{multicols}
\begin{equation}\label{eq:normalized}
\int_{-(\frac{h}{2}+h\frac{l}{d})+\mu_0}^{(\frac{h}{2}+h\frac{l}{d})+\mu_0}f(x){\rm d} x
=\alpha
\cdot
\bigg(
\int_{\frac{h}{2}+\mu_0}^{(\frac{h}{2}+h\frac{l}{d})+\mu_0}\Big[h-\frac{d}{l}(x-\mu_0-\frac{h}{2})\Big] {\rm d} x+
\int_{-\frac{h}{2}+\mu_0}^{\frac{h}{2}+\mu_0}h\ {\rm d} x+ \int_{-(\frac{h}{2}+h\frac{l}{d})+\mu_0}^{-\frac{h}{2}+\mu_0}\Big[h+\frac{d}{l}(x-\mu_0+\frac{h}{2})\Big] {\rm d} x
\bigg)
=1
\end{equation}
\begin{multicols}{2}
From Eq.~(\ref{eq:normalized}), we can get the normalization coefficient.
\begin{equation}
\alpha=1/{h^2(\frac{l}{d}+1)}
\end{equation}
It is worth to remind that if without the slit, emission angle of the line source is $2\pi$.

Curve of the beam intensity distribution, solid black line with shape of trapezoid(like a dam), is shown in Fig.~\ref{pic:manydistribution}. When $\sigma_0$ is very small, the experimental data distribution is close to the beam intensity distribution(dashed line in Fig.~\ref{pic:manydistribution}). Otherwise, when $\sigma_0$ is very large,  experimental data distribution do not reflect the beam intensity distribution but just reflect detector resolution function $r(x'-x)$ itself(dash-dotted line in Fig.~\ref{pic:manydistribution}).

\begin{center}
\includegraphics[width=80mm]{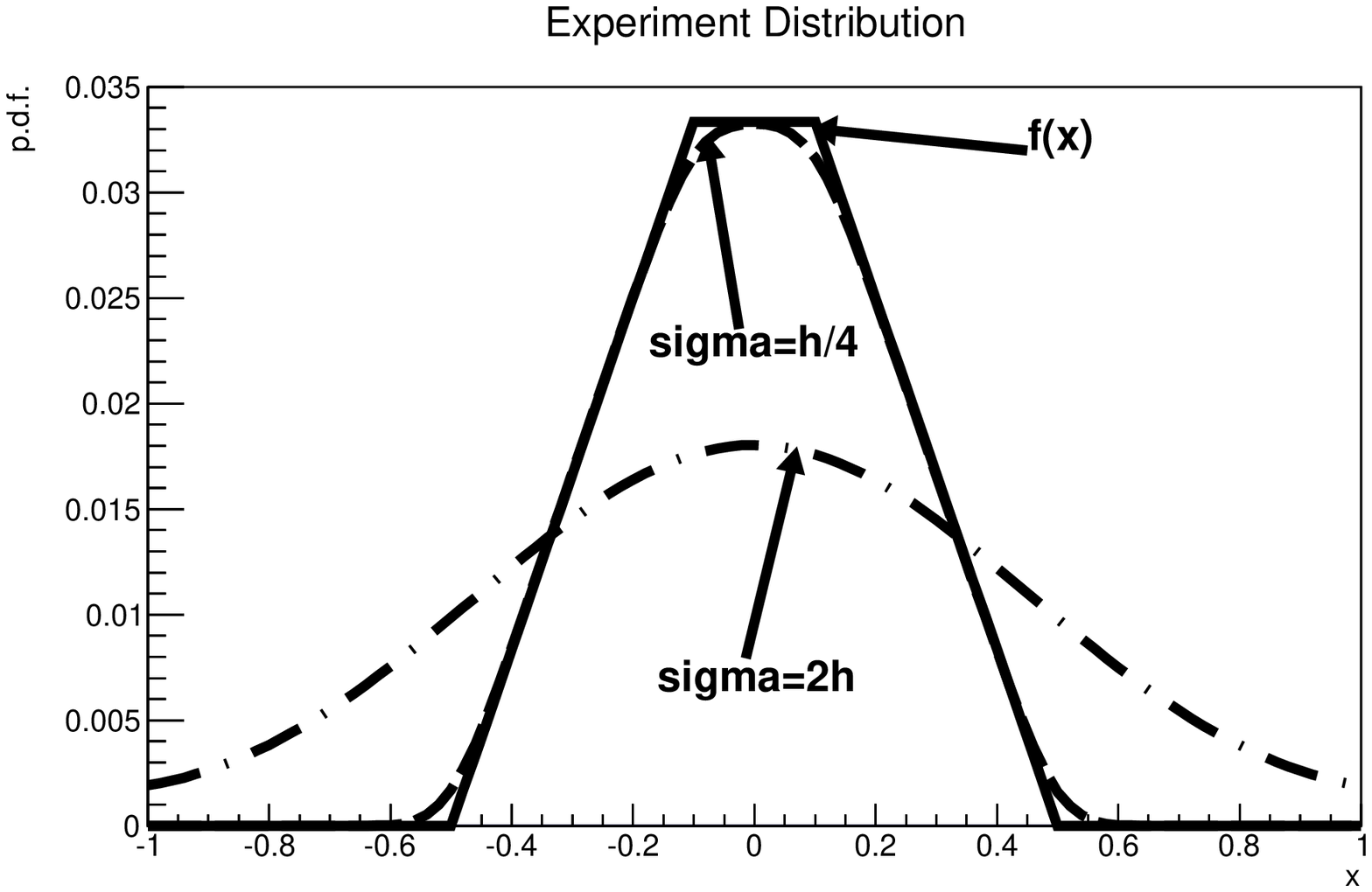}
\figcaption{\label{pic:manydistribution}Simulation study: the experimental data distributions as a function of the same collimator but different resolution of the detector. }
\end{center}

In other words, if the detector resolution is much larger than the slit width $h$, $g(x')$ is effectively approximated by $r(x'-x)$ and the collimator effect can be ignored, so an extremely narrow slit can be used as a collimator to verify these improved methods({Sec.~\ref{sec:validation}}).

\begin{center}
\includegraphics[width=80mm]{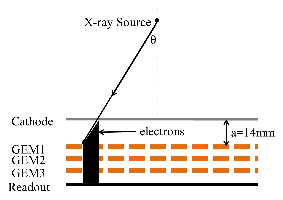}
\figcaption{\label{pic:angle}The schematic view of the incident angle affection.}
\end{center}

Attention should be paid to one more affecting factor of oblique incidence shown in Fig.~\ref{pic:angle}. Because of the incident angle $\angle\theta$, the quantity of the electrons on the top of GEM1 follows the uniformity distribution over the domain of $a\tan\theta$($a$ is the space of the drift area of the GEM detector) instead of a point. So, all of the measurement results of resolution in this paper should remove the incident angle affection by subtracting $a\tan\theta$.

In this method, the beam intensity distribution is known, the resolution function follows the normal distribution, of which the standard deviation $\sigma_0$ can be treated as an undetermined coefficient. When we measure the resolution in experiment, the convolution function(Eq.~(\ref{eq:bigconvolution})) is used to fit the experimental data distribution\cite{lab5} to obtain the resolution of the detector. The measurement result on GEM detector using this method is determined to be $\sigma_0=65.0\mu{}m$.

\end{multicols}
\begin{equation}\label{eq:bigconvolution}
g(x')=\alpha \cdot \left\{ \begin{array}{ll}
\int_{\frac{h}{2}+\mu_0}^{(\frac{h}{2}+h\frac{l}{d})+\mu_0}\frac{1}{\sqrt{2\pi}{\sigma_0}}e^{-\frac{(x'-x)^2}{2{\sigma_0}^2}}\cdot\big[h-\frac{d}{l}(x-\mu_0-\frac{h}{2})\big] {\rm d} x
&x\in\big[\frac{h}{2}+\mu_0, (\frac{h}{2}+h\frac{l}{d})+\mu_0\big]
\vspace{3mm}\\
\int_{-\frac{h}{2}+\mu_0}^{\frac{h}{2}+\mu_0}\frac{1}{\sqrt{2\pi}{\sigma_0}}e^{-\frac{(x'-x)^2}{2{\sigma_0}^2}}\cdot{}h\ {\rm d} x
&x\in\big(-\frac{h}{2}+\mu_0, \frac{h}{2}+\mu_0\big)
\vspace{3mm}\\
\int_{-(\frac{h}{2}+h\frac{l}{d})+\mu_0}^{-\frac{h}{2}+\mu_0}\frac{1}{\sqrt{2\pi}{\sigma_0}}e^{-\frac{(x'-x)^2}{2{\sigma_0}^2}}\cdot\big[h+\frac{d}{l}(x-\mu_0+\frac{h}{2})\big] {\rm d} x
&x\in\big[-(\frac{h}{2}+h\frac{l}{d})+\mu_0, -\frac{h}{2}+\mu_0\big]
\vspace{3mm}\\
\end{array}\right.
\end{equation}
The normalization coefficient
$\alpha=1/{h^2(\frac{l}{d}+1)}$
\begin{multicols}{2}

\subsection{Improved method~2: using a blade as the collimator with deconvolution and convolution fit}
{\label{sec:method2}}

In a 2-D image obtained from an imaging detector, it is reasonable that the sharper the image edge is, the more precipitous its projection histogram is. As shown in Fig.~\ref{pic:method2frame}, the beam intensity distribution of the edge is a step distribution if a blade is used to cut the beam. The beam intensity above the blade nearly follows the uniform distribution. The beam intensity below the blade is zero. As the resolution function is a Gaussian distribution, the experimental distribution is superposition of the step distribution and Gaussian distribution, which is in fact the cumulative Gaussian distribution\cite{lab6}.

\begin{equation}\label{eq:convolution2}
g(x')=\int_{x_{min}}^{x'}r(t){\rm d} t
\end{equation}

The resolution function can be obtained by derivation of $x$ on Eq.~(\ref{eq:convolution2}). $\sigma_0$ of the cumulative Gaussian distribution is the resolution. The process of solving the derivation(gradient) is that of deconvolution too. The measurement frame on GEM detector using this method is shown in Fig.~\ref{pic:method2frame} and the result is $\sigma_0=71.3\mu{}m$(Fig.~\ref{pic:method2result}). It is a little larger than the result of the first improved method, because the beam lines are not absolutely parallel.

\begin{center}
\includegraphics[width=80mm]{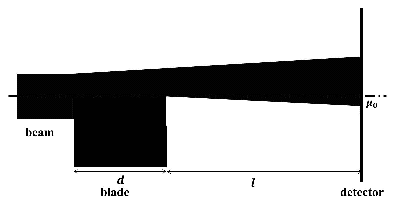}
\figcaption{\label{pic:method2frame}The schematic view of using blade as the collimator. $d=2cm$, $l=4cm$.}
\end{center}

\begin{center}
\includegraphics[width=80mm]{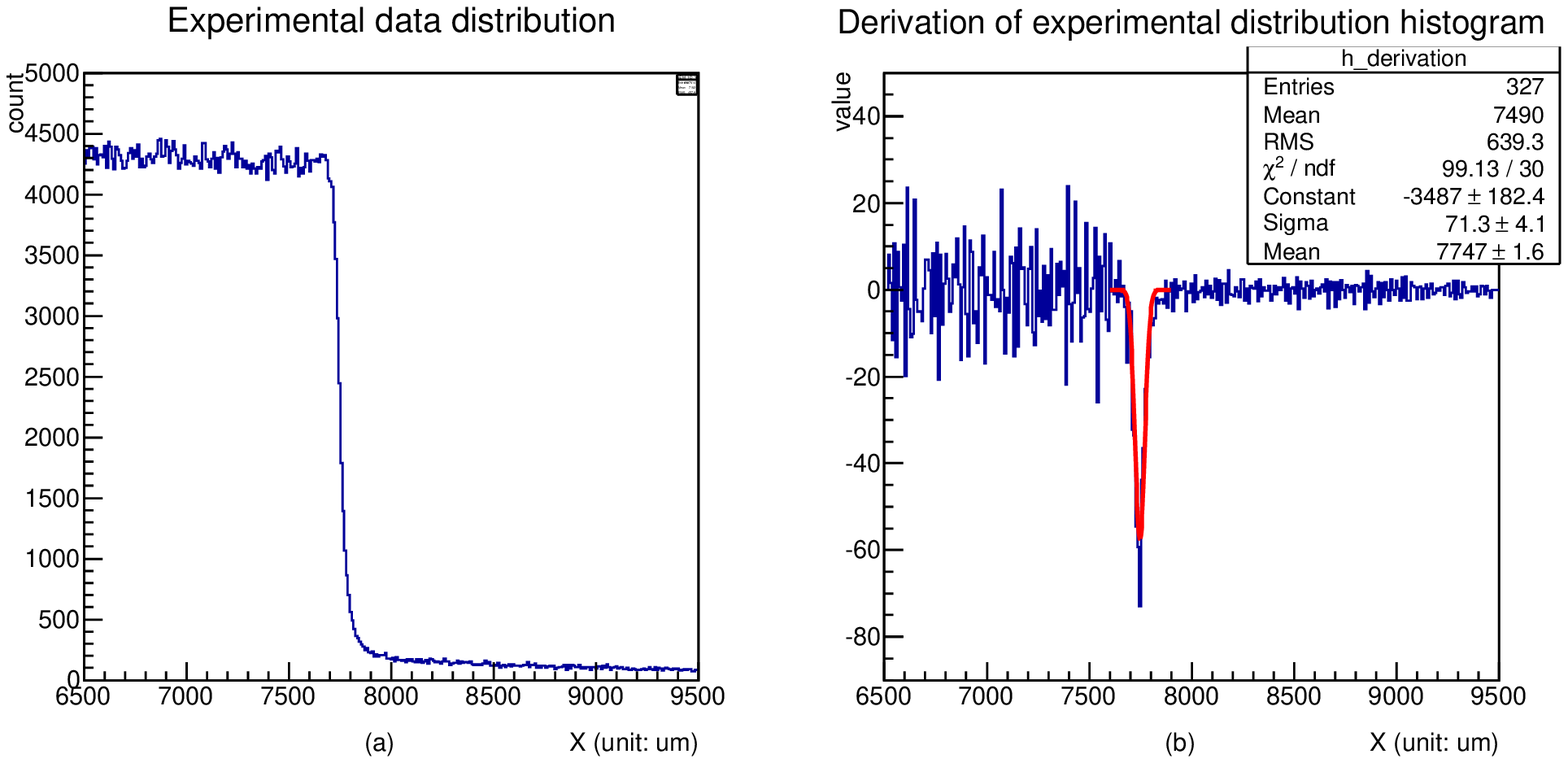}
\figcaption{\label{pic:method2result}Measurement result. (a) is the experimental data distribution; (b) is the derivation of the experimental data distribution.}
\end{center}

It is worthy to note that this method requires large statistics, otherwise, if there is no sufficient statistics per bin, taking the Gaussian distribution for an example, the derivation distribution of the same Gaussian p.d.f for different bin width will be different, as shown in Fig.~\ref{pic:zhendang}; which well yield different $\sigma_0$ for Gaussian p.d.f. In our case, the total statistic is $2\times10^6$, while the bin width is $10\mu{}m$.

\end{multicols}
\begin{center}
\includegraphics[width=150mm]{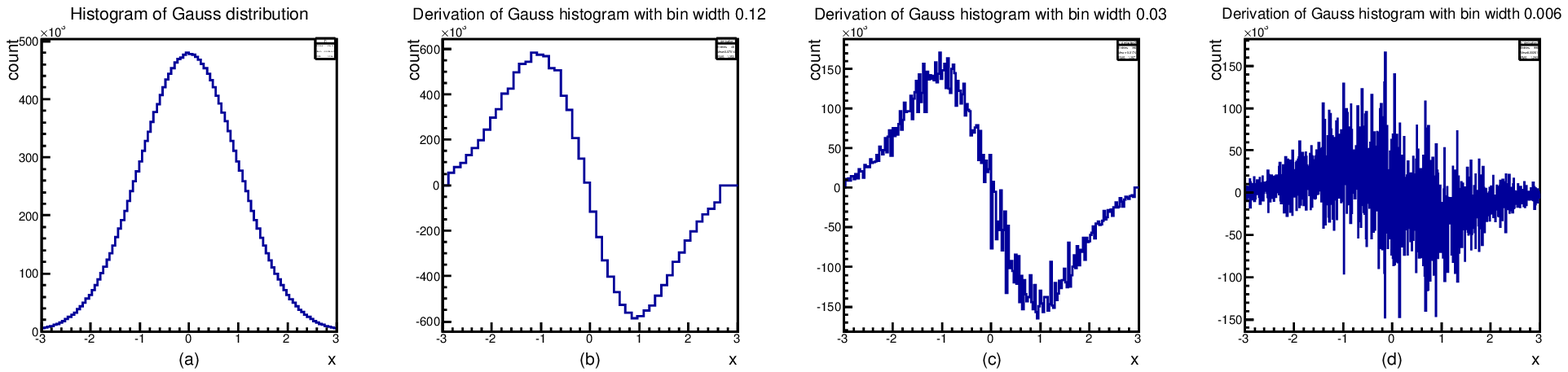}
\figcaption{\label{pic:zhendang}Limited statistics in a bin of a histogram for a Gaussian distribution will yield different derivation distributions. (a) is a Gaussian distribution; (b), (c), (d) are derivation of the Gaussian distribution with different bin-width.}
\end{center}
\begin{multicols}{2}

To avoid this kind of drawback, we improved this method, measurement architecture of which is the same to the deconvolution method but with a different data processing. As described above, the edge of an image follows cumulative Gaussian distribution. Indeed, the standard deviation of this Gaussian distribution is the resolution $\sigma_0$. So the cumulative Gaussian distribution with $\sigma_0$ as the undetermined coefficient can be used to fit the experimental data distribution. In this way, the resolution is determined to be $\sigma_0=63.3\mu{}m$ (Fig.~\ref{pic:cumulative}).

\begin{center}
\includegraphics[width=80mm]{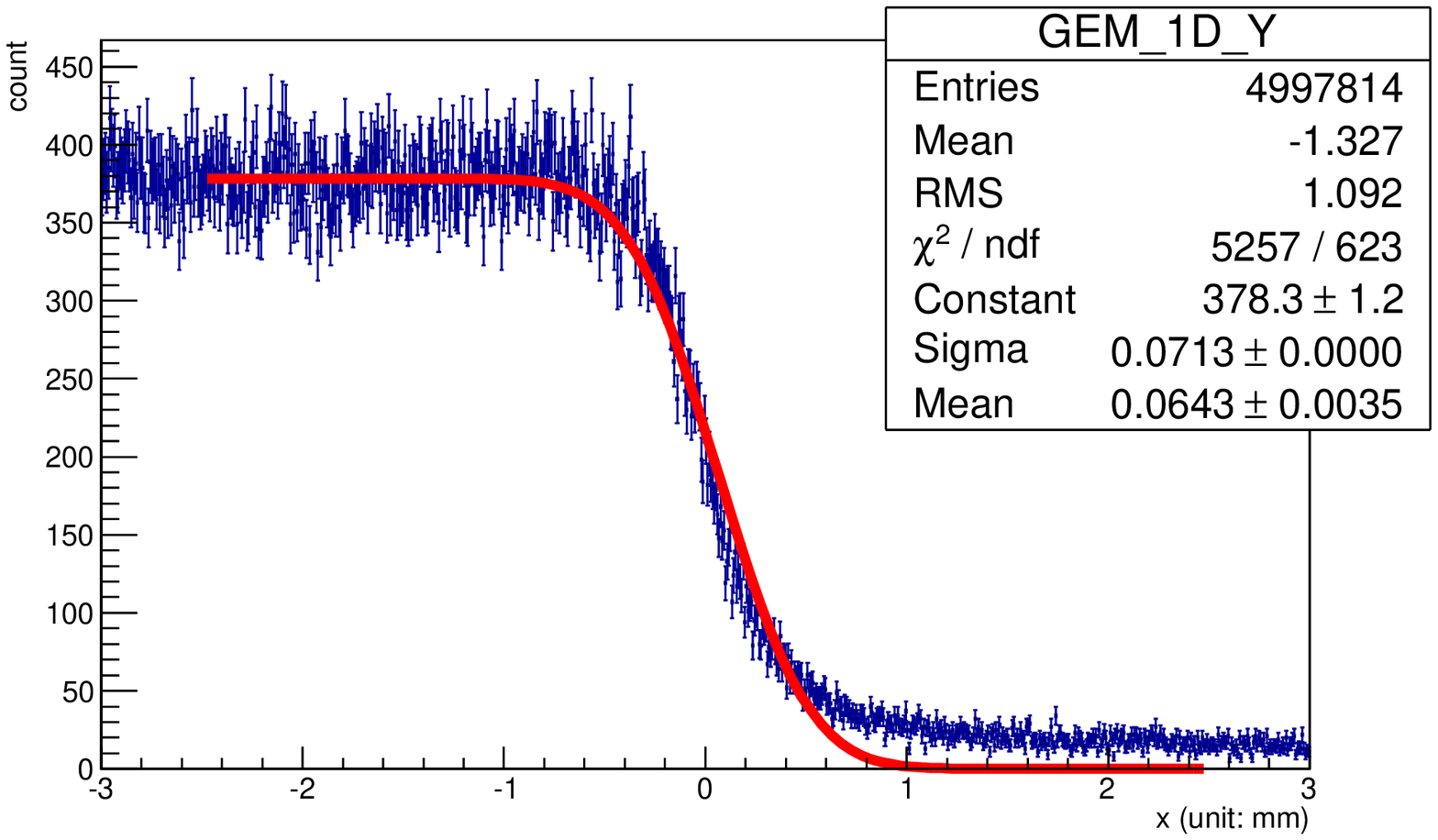}
\figcaption{\label{pic:cumulative}The experimental distribution using a blade as the collimator.}
\end{center}

\section{Verification the improved methods by using a extremely narrow slit as the collimator}\label{sec:validation}

As mentioned in {Sec.~\ref{sec:limitation}}, if the width of the collimator is narrow enough, less than tenth of the resolution(FWHM), the common method is applicable. In this method, an extremely narrow slit is used as the collimator. The measurement architecture is shown in Fig.~\ref{pic:slit}, where $h=0.01mm$, $l=40mm$ and $d=20mm$. Because the slit is so narrow that can it allow a very little part of beams to go through. High-intensity beams are required to get sufficient statistics. It is almost impossible to use the X-ray tube as the beam source in this situation. The measurement has been done at BSRF which can provide enough high intensity beams.

\begin{center}
\includegraphics[width=80mm]{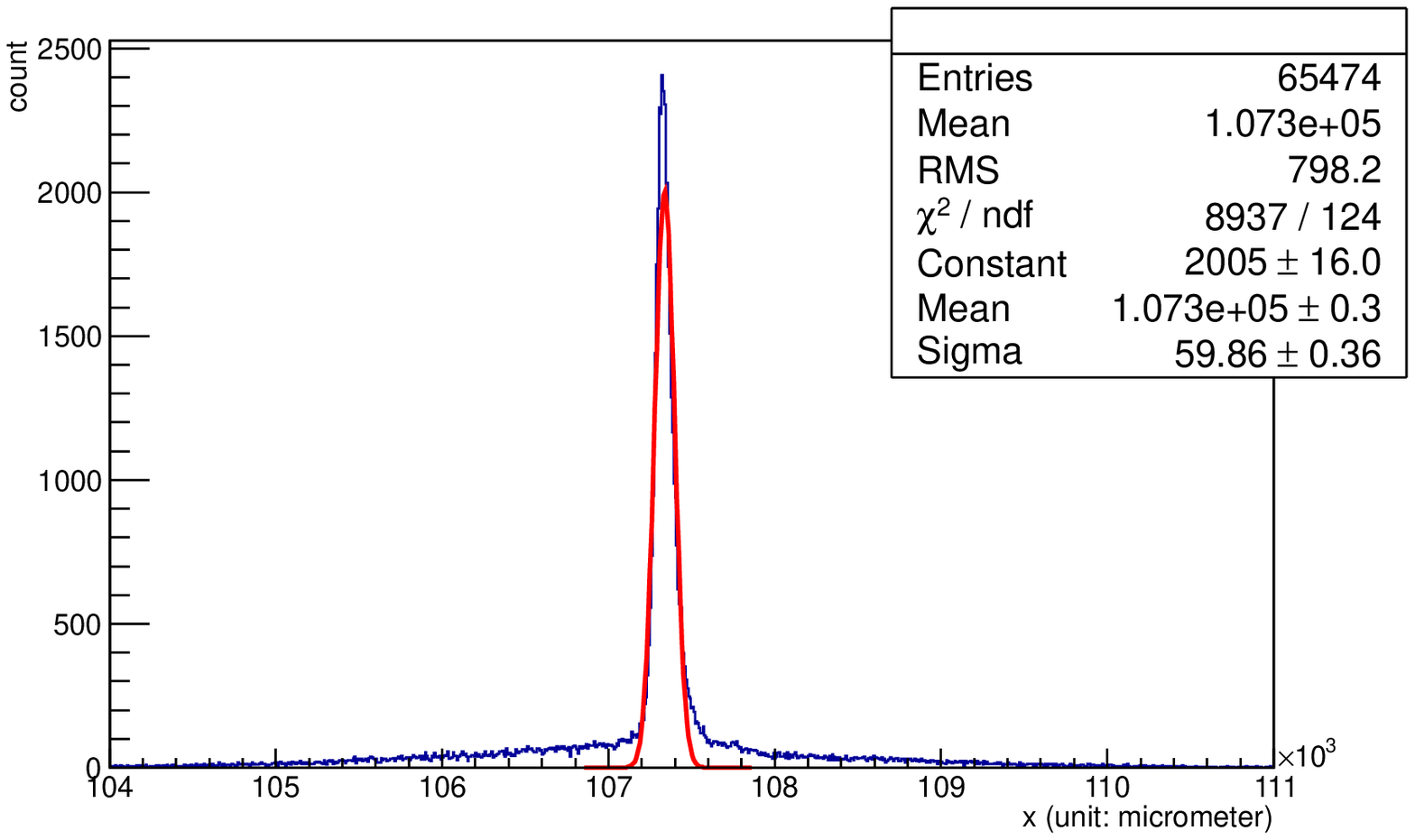}
\figcaption{\label{pic:method4shiyan}The measurement result using extremely narrow slit as the collimator. }
\end{center}

The result is shown in Fig.~\ref{pic:method4shiyan}. In this way, the resolution is $\sigma_0=59.9\mu{}m$. The result strongly validates the consistence with those of the improved methods.

\section{Measurement using the Rayleigh criterion}

The Rayleigh criterion is the generally accepted criterion for the minimum resolvable detail when the first diffraction minimum of the image of one source point coincides with the maximum of another. That is to say, the intensity of the saddle between two peaks of the points is $81\%$ the intensity of each peak\cite{lab7}.

\begin{center}
\includegraphics[width=80mm]{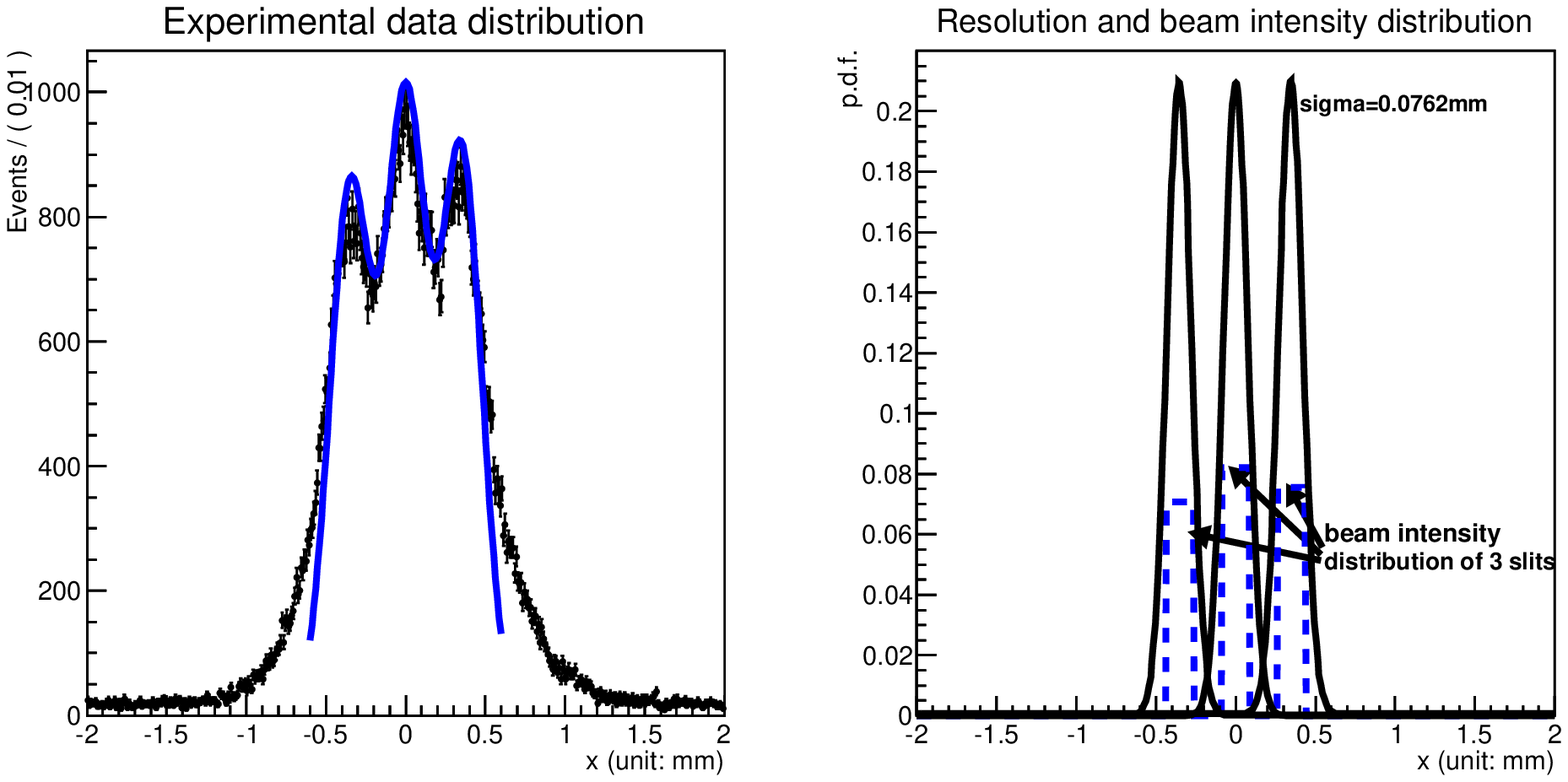}
\figcaption{\label{pic:rayleigh}The measurement result using the collimator with 3 slits. Left: the measurement data distribution; Right: the beam intensity distribution and the resolution function.}
\end{center}

To use the Rayleigh criterion, we have measured the distribution using a collimator, which has 3 slits with $0.32mm$ in pitch. Each slit is $0.20mm$ wide. The measurement result is shown in Fig.~\ref{pic:rayleigh}. As the width of each slit can not be ignored, the minimum resolvable pitch is $0.32mm$(Fig.~\ref{pic:rayleigh} left). After removing the effects of width of each slit by deconvolution fit, the resolution is $\sigma_0=76.2\mu{}m$. It is a little larger than the results of the first and second improved methods mentioned above. That is because of the definition of the Rayleigh criterion. If the considered function is the normal distribution, according to the Rayleigh criterion, when the intensity of the saddle between two peaks of the points is $81\%$ the intensity of each peak, the distance center to center of the two peaks is $2.69\sigma_0$. Yet, the distance of the first and second improved methods is the FWHM, which is $2.355\sigma_0$. So the resolution using the Rayleigh criterion is $1.142$ ($1.142=\frac{2.690}{2.355}$) times larger than those of the first and second improved methods.

\section{Summary and conclusion}\label{sec:conclusion}

The imaging detector's resolution can be limited by diffraction causing blurring of the image. By careful study of the relationship between the spatial resolution and the collimator, we give the scope of application of the common method and improve the method. The measurement using the improved methods gives more precise spatial resolution. Experimental validation has been done on the GEM detector as shown in Tab~\ref{tab:conclusion}.

\begin{center}
\tabcaption{\label{tab:conclusion}Summary of results of the measuring methods.}
\footnotesize
\begin{tabular*}{80mm}{c@{\extracolsep{\fill}}cc}
\toprule Method\#{}&$\sigma_0$ with incident angle correction\\
\hline
Common method&1314.9 $\mu{}m$\\
Method~1&65.0 $\mu{}m$\\
Method~2--1&71.3 $\mu{}m$\\
Method~2--2&63.3 $\mu{}m$\\
Extremely narrow slit&59.9 $\mu{}m$\\
Using the Rayleigh Criterion&66.7 $\mu{}m ^*$\\
\bottomrule
*~$66.7=76.2/1.142$.
\end{tabular*}
\end{center}

The measurement is an exploration of the measurement of spatial resolution and a general reference for that of other imaging detectors. Further study taking more factors in to account is in progress.

\end{multicols}
\vspace{-1mm}
\centerline{\rule{80mm}{0.1pt}}
\vspace{2mm}
\begin{multicols}{2}

\end{multicols}

\clearpage


\begin{thebibliography}{90}

\bibitem{lab1} XIE Y G, CHEN C et al. Particle detector and data acquisition(in chinese), Beijing: Science Press, 2003. 95---104.

\bibitem{lab2}Tommaso Lari. Nucl. Instr. and Meth. A, 2001, Vol.~465, 112---114

\bibitem{lab3}DONG J, XIE Y G, CHEN Y B et al. Chines Physics C, 2007, Vol.~31, NO.~7: 664---668

\bibitem{lab4}LIU B, DONG J, LU X Y et al. Acta Physica Sinica, 2010, Vol.~59: 6029---6035

\bibitem{lab5} W. Verkerke, D. kirkby. RooFit Users Manual v2.91, 2008.

\bibitem{lab6} ZHU Y S. Probability and statistics in experimental physics(in Chinese). Beijing: Science Press, Version 2, 2006. 47---51, 148---158

\bibitem{lab7} YAO Q J. Optics Tutorial(in Chinese). Beijing: Higher Education Press, Version 3, 2002. 287---290

\end{thebibliography}
\end{document}